\documentclass[11pt]{article}

\newcommand{\mb}{\mathbf}
\newcommand{\bs}{\boldsymbol}

\usepackage{natbib}
\usepackage{graphicx}

\usepackage{soul}
\usepackage{natbib}
\setcitestyle{authoryear,open={(},close={)}}

\usepackage{hyperref}
\usepackage{color}
\usepackage{multirow}

\usepackage{graphicx}
\usepackage{amsmath}
\usepackage{bbm}
\usepackage{comment}
\usepackage{setspace}
\newcommand{\blind}{1}

\begin{document}
\def\spacingset#1{\renewcommand{\baselinestretch}%
	{#1}\small\normalsize} \spacingset{1}

\if1\blind
{
	\title{\bf Sensitivity Analysis of Wind Energy Resources with Bayesian non-Gaussian and nonstationary Functional ANOVA}
	\author{Jiachen Zhang\\
	Department of Applied and Computational Mathematics and Statistics,\\
		University of Notre Dame (USA)\\
		and \\
		Paola Crippa\\
		Department of Civid and Environmental Engineering and Geoscience,\\
		University of Notre Dame (USA)\\
		and\\
		Marc G. Genton\\
		Statistics Program,\\
		King Abdullah University of Science and Technology (Saudi Arabia)\\
		and\\
		Stefano Castruccio\thanks{scastruc@nd.edu}\hspace{.2cm}\\
		Department of Applied and Computational Mathematics and Statistics,\\
		University of Notre Dame (USA)\\
		\\
	}
	\maketitle
} \fi

\newpage

\begin{abstract}
The transition from non-renewable to renewable energies represents a global societal challenge, and developing a sustainable energy portfolio is an especially daunting task for developing countries where little to no information is available regarding the abundance of renewable resources such as wind. Weather model simulations are key to obtain such information when observational data are scarce and sparse over a country as large and geographically diverse as Saudi Arabia. However, output from such models is uncertain, as it depends on inputs such as the parametrization of the physical processes and the spatial resolution of the simulated domain. In such situations, a sensitivity analysis must be performed and the input may have a spatially heterogeneous influence of wind. In this work, we propose a latent Gaussian functional analysis of variance (ANOVA) model that relies on a nonstationary Gaussian Markov random field approximation of a continuous latent process. The proposed approach is able to capture the local sensitivity of Gaussian and non-Gaussian wind characteristics such as speed and threshold exceedances over a large simulation domain, and a continuous underlying process also allows us to assess the effect of different spatial resolutions. Our results indicate that (1) the non-local planetary boundary layer scheme and high spatial resolution are both instrumental in capturing wind speed and energy (especially over complex mountainous terrain), and (2) the impact of planetary boundary layer scheme and resolution on Saudi Arabia's planned wind farms is small (at most 1.4\%). Thus, our results lend support for the construction of these wind farms in the next decade. 
\end{abstract}

\noindent%
\vfill

\spacingset{2} 

\addtolength{\textheight}{.5in}%

\section{Introduction}

Global fossil fuel consumption has increased eight-fold since 1950 and remains the major contributor to global climate change and atmospheric air pollution \citep{yil18, ren21}. Given that oil is a non-renewable energy resource, the rapid depletion of its reserves is a problem that humanity will face in the foreseeable future, prompting the development of alternative, renewable sources of energy. Wind energy, the focus of this work, grew significantly in recent decades and has already contributed to the reduction of greenhouse gas emissions
and local air pollution. According to recent estimates, in 2019, 5\% of the world's electricity was generated by wind, with the largest share coming from the United States and China \citep{renew,ren21}. Further, although European countries have a smaller absolute installed wind capacity, their share in the energy portfolio is one of the highest and in Denmark, wind is by far the most widespread form of energy (56.3\%). 

Despite the growth of wind energy in many countries, in areas such as the Middle East and North Africa (MENA) the proportion of power generated by this renewable energy is among the lowest worldwide \citep{moh21}. Saudi Arabia, in particular, is almost exclusively reliant on its abundant oil reserves for its internal energy demand; it has only recently outlined plans to diversify its energy portfolio using renewable energy sources. Given the country’s latitude and climate, solar energy is expected to be the primary form of renewable energy \citep{agh20}, but the recently proposed `Vision 2030' plan aims to generate 16 GW of wind energy, which will place the country in a leading position for wind energy generation \citep{nur17,nre18}. Unlike solar energy, which can only be harnessed during daytime, wind is always available and often peaks during nighttime, thus providing a promising opportunity to integrate complementary renewable energy resources and hence provide a continuous and reliable supply to the grid.

A comprehensive assessment of wind energy resources in a country as large as Saudi Arabia cannot be informed solely by observational records, which are very sparse in space, limited in time and challenging to retrieve. Instead, assessments must be integrated with climate model simulations, which provide spatially resolved, dynamic and physically consistent information on wind speed. Over the past few years, considerable progress has been made in this regard. The initial efforts were directed towards analyzing publicly available simulations from either global models \citep{jeo18,jeo19} or regional simulations \citep{che18}, such as the MENA COordinated Regional climate Downscaling EXperiment (MENA CORDEX, \cite{jon11}). More recently new and more detailed assessments have been performed by analyzing \textit{ad hoc} high resolution regional simulations \citep{tag19,gia20,zhang21,cri21} from a state-of-the-art model: the Weather Research and Forecasting (WRF, \cite{ska19}). 

Numerical models can provide a comprehensive, spatially resolved assessment of wind over a large country, but each simulation depends on several inputs, including the parameterizations for physical processes (especially near the surface), boundary conditions, as well as spatial and temporal resolution. Therefore any assessment must be performed using a collection (\textit{ensemble}) of simulations, with each ensemble member representing a different choice of the aforementioned input. In order to provide a reliable uncertainty quantification of wind energy, it is therefore crucial to understand if, to what extent, and where the final wind energy estimates depend on these model choices. 

From a methodological point of view, sensitivity analysis of a variable dependent on factors with multiple levels is one of the oldest and most established problems in statistics, long before the formulation of numerical simulations. The significance of a factor is assessed by comparing its variability to the measurement error (analysis of variance, ANOVA). In its original formulation, ANOVA is aimed at sensitivity analysis of a single variable and a finite number of factors with different levels being observed independently. More flexible models have been proposed to perform ANOVA on more complex data structures. In this work we focus on functional ANOVA (FANOVA), a method developed to perform sensitivity analysis on spatial, temporal, or even spatio-temporal process \citep{stone97}. FANOVA has been used in many fields such as public health \citep{hua00,zha19,ull13}, chemistry \citep{saa11} and geoscience \citep{sain,sain2,sun12,qu21}, and the majority of the applications focused on time series. \cite{stone97} provided a comprehensive review of FANOVA. More recently \cite{zha09} proposed a smoothed ANOVA model using a Bayesian framework that treated space as a factor in the model for multivariate observations with an areal spatial structure. \cite{sain} proposed a Bayesian framework for spatial FANOVA to perform sensitivity analysis of present and future regional climate simulations. Their work laid the foundation for a framework to perform the local sensitivity analysis of climate simulations, but has several limitations. First, the proposed model assumes Gaussian data, thereby limiting the analysis to continuous variables at a sufficiently high level of temporal aggregation. Second, the spatial dependence structure was restricted to stationary isotropic models. Although this is a useful simplifying assumption, it is unrealistic for large simulation domains \citep{cres99, mik08, yue10}. Finally, the assumption of a discrete spatial model without an underlying continuous process would not allow one to perform FANOVA to ensemble members with differing spatial resolution. 

In this work, we propose a new FANOVA model that is able to overcome the aforementioned methodological limitations. The key idea is that the latent process in the model can be regarded as the solution of a stochastic partial differential equation (SPDE). More specifically:
\begin{description}
 
\item [1.] The proposed model relies on the flexible class of latent Gaussian models: the marginal distribution is assumed to be a member of the exponential family, with the expected values being spatially dependent through an appropriate link function and one or more latent Gaussian fields. Bayesian inference for this class of latent Gaussian models can be performed by deterministic approximation of the integrals in the posterior distribution using the integrated Laplace approximation (INLA) method \citep{rue}.

\item [2.] The definition of a spatial model through a solution of an SPDE allows one to naturally generalize FANOVA to a nonstationary setting by generalizing the differential operator and assuming its non-homogeneity in space while still allowing a theoretically valid model. The SPDE chosen allows for an `explicit link' between the SPDE solution and the Gaussian Markov Random Field \citep{lindgren}. This allows likelihood evaluations with a sparse precision matrix, and hence fast and affordable inference for very large spatial data. 
 
\item [3.] The use of a FANOVA with a latent Gaussian field assumes an  underling continuous spatial model, which naturally accounts for datasets with different resolutions. 
\end{description}

From a methodological standpoint, our work generalizes the FANOVA approach in \cite{yue19}, by allowing a flexible class of nonstationary models (while also exploring others in the supplementary material) for the latent Gaussian field which, as the extensive simulation study in this work will show, is more flexible in capturing the spatial patterns, and also explores space-time interaction models.

While the proposed model is motivated by assessing the robustness of Saudi Arabia's current plan for wind energy installation, its scope is more general as it can be applied to any model ensemble resolved in space and time where one seeks a local sensitivity analysis with respect to a controlled number of factors.

This work proceeds as follows. In Section \ref{sec:data} we introduce the data set of numerical simulations used for this study. Sections \ref{sec:model} and \ref{sec:infer} describe the proposed statistical methodology and inference, respectively. Section \ref{sec:sim} validates the statistical model with a simulation study. In Section \ref{sec:app}, we apply the proposed method to the WRF simulated data of wind speed to assess wind energy sensitivity across Saudi Arabia. Section \ref{sec:con} provides the conclusions of this study. The code for this work is available at the following GitHub repository: github.com/Env-an-Stat-group/21.Zhang.unpublished.public.

\section{Data Description}\label{sec:data}

\begin{figure}[ht!]
\includegraphics[scale=0.57]{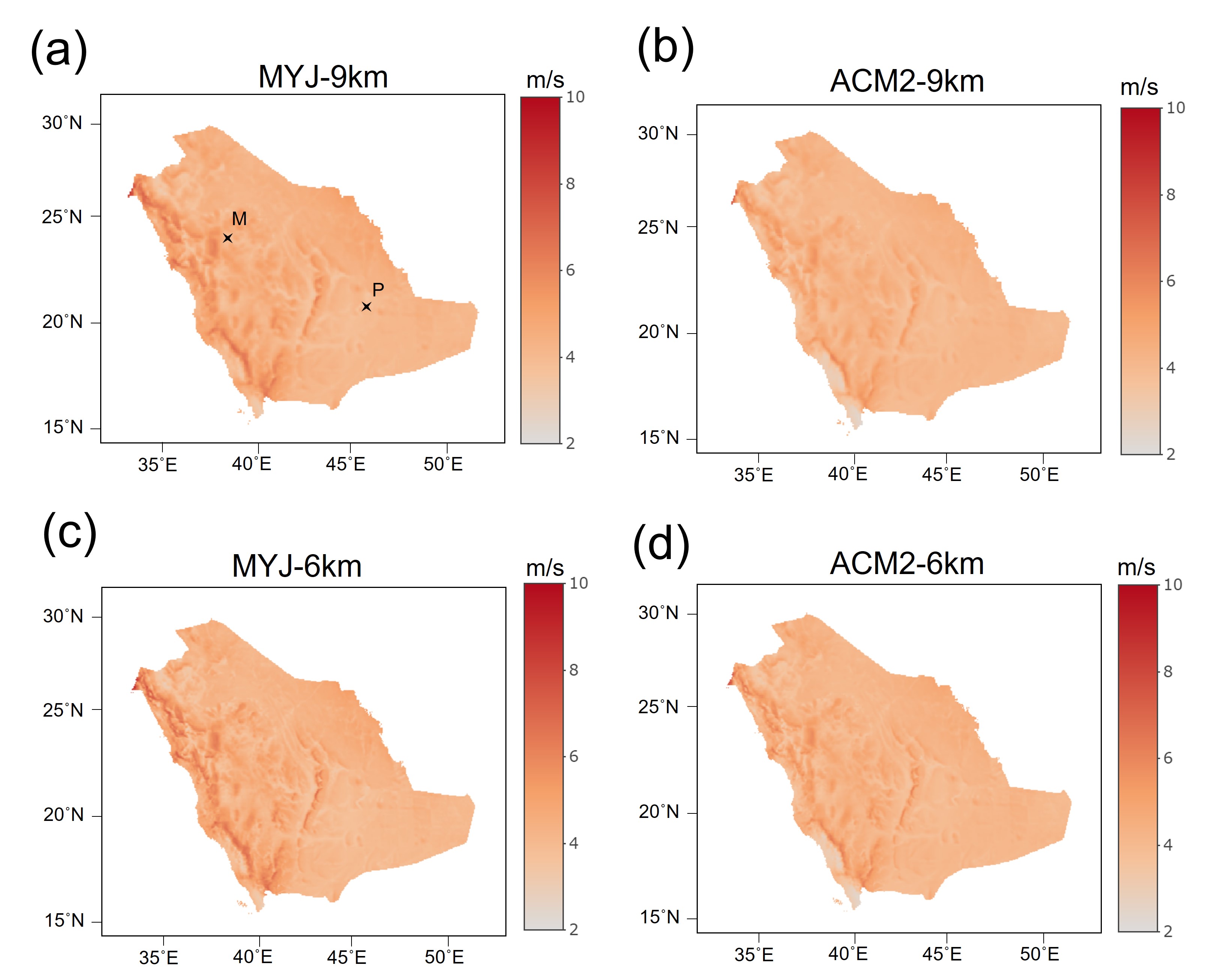}
\caption{Wind speed averaged across the years 2013-2016 simulated using the Weather and Research Forecasting (WRF) model in (a) MYJ-9km, (b) ACM2-9km, (c) MYJ-6km, and (d) ACM2-6km.}
\label{fig:data}
\end{figure}

This study relies on an ensemble of high resolution WRF simulations of the Arabian Peninsula during the years 2013-2016 \citep{gia20}. For the ease of understanding, we show a topographical map of Saudi Arabia, along with the names of the regions that will be used throughout this work in Figure S1. The ensemble has been designed to explore model's sensitivity towards different spatial resolutions and Planetary Boundary Layer (PBL) schemes. The simulated domain comprises 339 $\times$ 299 and 549 $\times$ 499 grids at a resolution of 9km and 6km, respectively. Wind speed is resolved on a vertical grid comprising 40 layers, which are more closely spaced near the surface and sparser closer to the boundary layer height. In this work we consider wind between 10 and 110m, which correspond to the heights of the majority of commercial wind turbines. The initial and boundary conditions used to drive WRF are obtained from high-resolution European Centre for Medium-Range Weather Forecast (HRES-ECMWF, \cite{rda}) model. The two PBL schemes adopted are: the Mellor–Yamada–Janjić (MYJ, \cite{myj}) and the asymmetric convective model (ACM2, \cite{acm2}). MYJ is a local scheme where the vertical diffusion occurs only between neighboring cells, while ACM2 is a non-local scheme, in which the diffusion also occurs between non-neighboring cells and counter-gradient fluxes. The spatial resolution of the simulation and PBL setup among the different runs are summarized in Table \ref{tab:sum}. Figure \ref{fig:data} shows the simulated domain and the wind speed averaged over the simulated time period for all four ensemble members. Discrepancies in the magnitude of the simulated wind speed are observed across different regions and runs, with higher resolution simulations being able to more accurately characterize the topography (and hence the wind speed) in mountainous regions, and MYJ resulting in higher winds at finer resolution (see panels (c) and (d)). A spatial sensitivity analysis is therefore necessary to quantify the differences across resolutions and PBL schemes, and motivates the development of the FANOVA model in the next section. 

\begin{table}
	\caption{Resolution and planetary boundary layer (PBL) setup of the four ensemble members.\label{tab:sum}}
	\centering
	\begin{tabular}{|l|l|l|l|}\hline 
		Run & Resolution & PBL  \\ \hline
		1 & 9km & MYJ \\ \hline
		2 & 9km & ACM2\\ \hline
		3 & 6km & ACM2\\ \hline 
		4 & 6km & MYJ\\ \hline  \end{tabular}
\end{table}

\begin{figure}[ht!]
\includegraphics[scale=0.46]{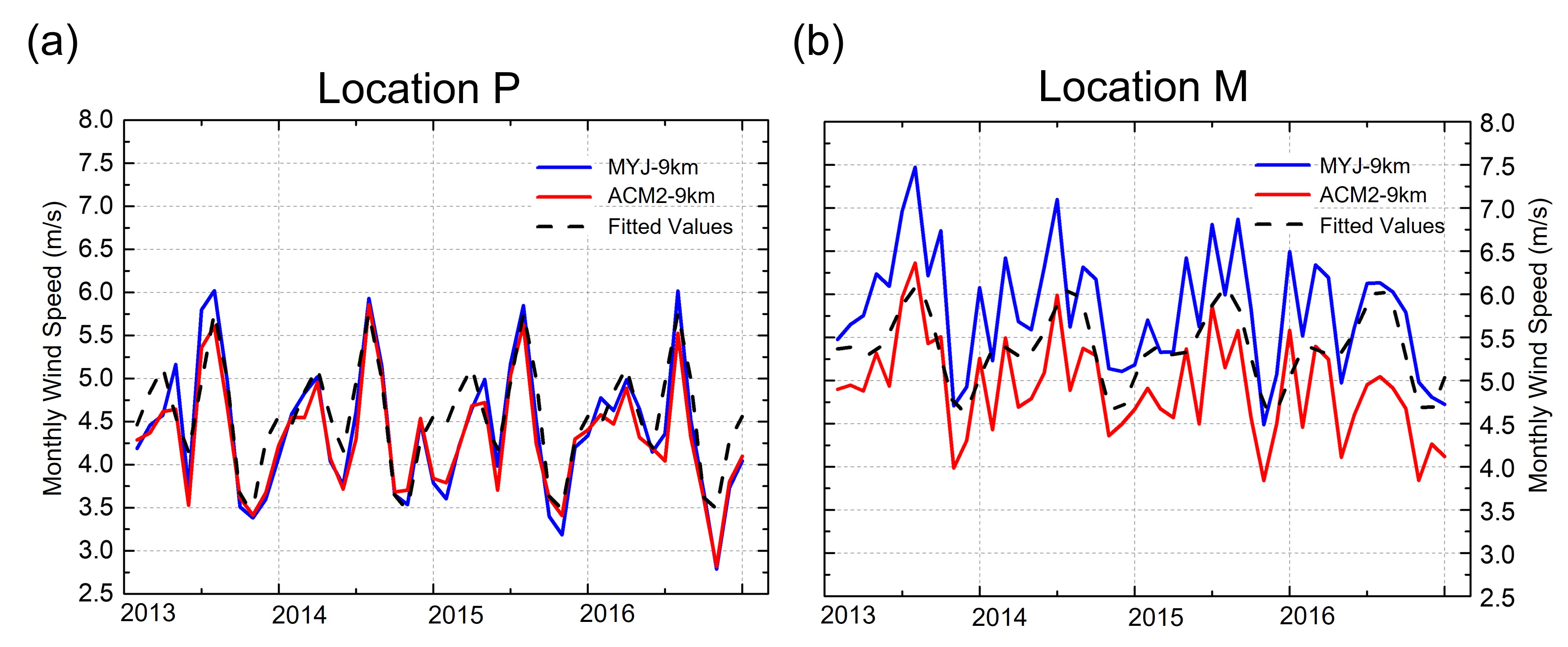}
\caption{MYJ-9km and ACM2-9km monthly wind speed and fitted value according to model \eqref{eq:times} at two locations P (Plain) and M (Mountain) indicated by the crosses in Figure \ref{fig:data}(a).}
\label{fig:wind}
\end{figure}

In Figure \ref{fig:wind}, the monthly wind speed at two selected locations, represented by crosses in Figure \ref{fig:data}(a), shows an interannual, seasonal behavior. Wind speeds are the highest during the summer and the lowest during winter at location P, a plain area within the Rub' al Khali region (see Figure S1 for an indication of Saudi Arabia's regions) and the opposite at location M in the Hijaz mountain region. The annual cycle observed is attributable to Saudi Arabia's location within the trade-wind belt of the Northern Hemisphere \citep{has15}. The strong northerly flow contributes to the high wind speeds during the summer. In addition, the southeastern wind from the Indian Ocean travels to the southeastern part of the Arabian Peninsula during the monsoon season, further enhancing the summer wind. However, during winter, the north wind traveling from Mediterranean to the Persian Gulf weakens. 

\begin{sloppypar}
Throughout this work, we denote with $N$ the total number of locations, and by $\bs{Y}_{ij}(t)=(Y_{ij}(\mb{s}_1, t), \ldots, Y_{ij}(\mb{s}_N, t))^\top$ the $N$-dimensional vector of the variable we are interested in (all aggregated at monthly level) at locations $\mb{s}_1, \ldots, \mb{s}_N$, for the $i$th PBL settings ($i=0, 1$ indicate MYJ and ACM2, respectively) the $j$th resolution ($j=0, 1$ indicate 9km and 6km, respectively), and the $t$-th month over the four year time span ($t=1, \ldots, 4 \times 12$).
\end{sloppypar}

\section{Model}\label{sec:model}

We propose a Bayesian spatio-temporal non-Gaussian FANOVA model with nonstationary dependence. Because the ensemble introduced in Section \ref{sec:data} has PBL and resolution as inputs we assume two predictors, however the model can be generalized to any number:

\begin{equation} \label{eq:baye}
 \begin{array}{rcl}
 \bs{Y}_{ij}(t) &\sim& h (\bs{\mu}_{ij}(t),\bs{\theta}_{\text{MRG}}), \\[7pt]
 \mb{g}(\bs{\mu}_{ij}(t)) & = & \bs{f}^{(T)}(t) + \bs{f}_{ij}^{(F)}+ i\bs{\beta}_{\text{PBL}}+j\bs{\beta}_{\text{RES}} +\bs{\epsilon}_{ij}(t),\\[7pt]
 
 \bs{f}^{(T)}(t) & = & \sum_{k=1}^K \left\{\bs{\zeta}_{k} \sin\left(\frac{2\pi k t}{\delta}\right)+\bs{\zeta}'_{k} \cos\left(\frac{2\pi k t}{\delta}\right)\right\},\\[7pt]
 \bs{f}^{(F)}_{ij} & = & \beta_{\text{ALT}}\bs{A}_{j},
 \end{array}
\end{equation}
where bold indicates spatial vectors, e.g., $\bs{\zeta}_{k}=(\zeta_{k;1},\ldots, \zeta_{k;N})^\top$ and similarly for all other vectors. Each element of the vector $\bs{Y}_{ij}(t)$ has a distribution from the exponential family $h(\cdot)$, with an expectation $\bs{\mu}_{ij}(t)$ and possibly other marginal parameters $\bs{\theta}_{\text{MRG}}$. The vector $\mb{g}(\cdot)$ consists of the same link function $g(\cdot)$ for every location, which depends on the distribution (e.g., for the Gaussian distribution we have the identity, whereas for the Bernoulli distribution we have a logit). We assume that the vector of expectations $\bs{\mu}_{ij}(t)$ is modeled using a location-specific $N$-dimensional vector of time effects $\bs{f}^{(T)}(t)$, a location-specific (possibly simulation-dependent) $N$-dimensional vector of time-invariant (fixed) effects $\bs{f}_{ij}^{(F)}$, two Gaussian $N$-dimensional spatial random vectors $\bs{\beta}_{\text{PBL}}, \bs{\beta}_{\text{RES}}$ which represent the local contribution of PBL and resolution, respectively. In the supplementary material we also consider a more general model with space-time interaction, which did yield very marginal improvements but was not implemented for the sake of simplicity and interpretability. Finally, $\bs{\epsilon}_{ij}(t)$ represent a Gaussian white noise independent of factors, time and space with variance $\sigma^2$. 

\begin{sloppypar}
The time effect $\bs{f}^{(T)}(t)$ is described using $K$ annual harmonics controlled at each location $\mb{s}_n$ by different parameters $\bs{\theta}_{\text{time},n}=\{\zeta_{k;n}, \zeta'_{k;n}, k=1,\ldots, K\}$, so that the total number of temporal parameters is $\bs{\theta}_{\text{time}}=\{\bs{\theta}_{\text{time},n}, n=1,\ldots, N\}$. In our case, the time invariant effect $\bs{f}_{ij}^{(F)}$ represents the contribution of the altitude, which is expected to be linear according to a parameter $\beta_{\text{ALT}}$, since wind is generally higher in mountainous regions with more complex terrain, see Figure S2 and associated diagnostics. In our application, the altitude map $\mb{A}_j$ depends on the resolution of the simulation, hence the subscript $j$ is used. In a standard two-way ANOVA, the two random effects $\bs{\beta}_{\text{PBL}}$ and $\bs{\beta}_{\text{RES}}$ will be independent in space, but in FANOVA they are assumed to be the realizations (independent in time) from a Gaussian random field: 
\end{sloppypar}
\begin{equation} \label{eq:fanova}
\begin{array}{rcl}
 \bs{\beta}_{\text{PBL}} & \sim &   \mathcal{N}(\mb{0}, \bs{\Sigma}(\bs{\theta_{\text{PBL}}})), \ \\[7pt] 
 \bs{\beta}_{\text{RES}} & \sim &  \mathcal{N}(\mb{0}, \bs{\Sigma}(\bs{\theta_{\text{RES}}})),
 \end{array}
\end{equation}
where $\bs{\theta_{\text{PBL}}}$ and $\bs{\theta_{\text{RES}}}$ are unknown parameters.

In this work we assume $\bs{\Sigma}(\bs{\theta}_{\ell})=(\bs{\Sigma}_{ij}(\bs{\theta}_{\ell}))_{ij}$  where $\ell \in \{\text{PBL}, \text{RES}\}$. Instead of providing an explicit parametrization of the covariance matrix through a covariance function, we use a fundamental result in spatial statistics that links a class of covariance functions to the solution of a SPDE. Specifically, we focus on the Mat\'ern function  
\begin{equation}\label{eq:matern}
\bs{\Sigma}_{ij}(\bs{\theta}_{\ell})= \frac{1}{\tau_{\ell} 2^{\nu_{\ell}-1}\Gamma(\nu_{\ell})}(\kappa_{\ell}\|\mb{s}_i-\mb{s}_j\|)^{\nu_{\ell}} K_{\nu_{\ell}}(\kappa_{\ell}\|\mb{s}_i-\mb{s}_j\|),
\end{equation}
where $K_\nu$ is the modified Bessel function of the second kind of order $\nu$, $\|\bs{s}_i-\bs{s}_j\|$ is the Euclidean distance between two generic locations $\bs{s}_i, \bs{s}_j\in \mathcal{R}^2$, and the parameter vector is $\bs{\theta}_{\ell}=(\tau_{\ell}, \kappa_{\ell},  \nu_{\ell})^\top$. The parameter $\tau_{\ell}$ controls the marginal precision and $\kappa_{\ell}$ controls the range of the spatial dependence: when we consider a distance $\sqrt{8\nu_\ell}/\kappa_\ell$, the spatial correlation is near 0.1 for all $\nu_\ell$ \citep{ste99}. Finally, $\nu_{\ell}$ controls the degree of smoothness of the process and is usually fixed because of poor identifiability.

A well known result \citep{wht54,wht63} stipulates that a Gaussian process with Mat\'ern covariance is the (unique) stationary solution of a fractional diffusion-reaction SPDE:
\begin{equation}\label{eq:spde}
(\kappa_{\ell}^2-\Delta)^{\nu_{\ell}/2+1/2}(\tau_{\ell}\bs{\beta}_{\ell}) = \mathcal{W}(\mb{s}),
\end{equation}
where $\Delta$ is the Laplacian operator, $\tau_\ell$ controls the variance and $\mathcal{W}(\mb{s})$ is a spatial Gaussian white noise with unit variance. 

A stationary model resulting from solving \eqref{eq:spde} is overly simplistic over a large simulated domain such as the one used in our application. However, one of the main advantages of the SPDE approach is that it can be used as a baseline to formulate nonstationary models that are automatically well defined, a task considerably more challenging to perform when relying on covariance-based models. Several options are available to achieve a nonstationarity, ranging from nested SPDEs \citep{bol11} to modification of the differential operator \citep{fug19}. Some recent works have proposed a nonstationary model based on a local deformation of the SPDE with a changing dependence structure across large geographical descriptors such as land and ocean in a Euclidean \citep{fug19} and global \citep{fug20,hu21} domain. Since in this work the geography of the problem does not suggest a natural partition of the domain, we instead rely on a basis decomposition approach, and assume that $\bs{\beta}_{\ell}$ is a solution of a generalization of \eqref{eq:spde} with varying range and precision:
\begin{equation}\label{eq:nonstat}
(\kappa_{\ell}^2(\mb{s})-\Delta)^{\nu_{\ell}/2+1/2}\{\tau_{\ell}(\mb{s})\bs{\beta}_{\ell}\} = \mathcal{W}(\mb{s}),
\end{equation}
where $\nu_{\ell}$ is fixed, and $\tau_{\ell}(\mb{s})$ and $\kappa_{\ell}^2(\mb{s})$ change in space according to some basis function:
\begin{equation} \label{eq:basis}
\begin{array}{rcl}
 \log(\tau_{\ell}(\mb{s})) &=& b_{0;\ell}^{(\tau)}(\mb{s})+\sum_{k=1}^p b_{k;\ell}^{(\tau)}(\mb{s})\theta^{(\tau)}_{k;\ell}, \\[7pt] 
  \log(\kappa_{\ell}(\mb{s})) &=& b_{0;\ell}^{(\kappa)}(\mb{s})+\sum_{k=1}^p b_{k;\ell}^{(\kappa)}(\mb{s})\theta^{(\kappa)}_{k;\ell},
 \end{array}
\end{equation}
\begin{sloppypar}
so that the total number of spatial parameters is $\bs{\theta}_{\text{space}}=\left\{b^{(\kappa)}_{k;\ell}, b^{(\tau)}_{k;\ell}, k=0,\ldots, p, \ell\in \{\text{PBL}, \text{RES}\}\right\}$. All parameters are assumed to have independent vague Gaussian priors with mean zero and variance equal to 1,000. 

In the supplementary material, we have also tested the barrier model \citep{bak19} to account for potentially abrupt changes between plains and mountains, and obtained far superior results with the choice of nonstarionary model in \eqref{eq:nonstat} and \eqref{eq:basis}.
\end{sloppypar}

\section{Inference}\label{sec:infer}
Inference is performed in two steps, in order to reduce the overall dimension of the parameter space at each step. First,  $\bs{f}^{(T)}(t)$ in \eqref{eq:baye} is estimated independently for each location to capture the annual periodicity of the (latent) Gaussian field. Second, inference on $\bs{\theta}_{\text{MRG}}, \mb{\beta}_{\text{ALT}}, \bs{\theta}_{\text{space}}$ is performed conditionally on the posterior mean of the parameters of $\bs{f}^{(T)}(t)$. A numerical study of this two-steps approach against joint space-time inference on a smaller subset is performed in the supplementary material, showing that the posteriors are similar for both approaches. 

\subsection{Step 1: Temporal Structure}

In the first step, we consider each location $\mb{s}_n$ independently, and a marginal time series version of \eqref{eq:baye} with no spatial and covariate effects: 

\begin{equation} \label{eq:times}
 \begin{array}{rcl}
 Y_{ij}(\mb{s}_n,t) &\sim& h (\mu_{ij}(\mb{s}_n,t),\bs{\theta}_{\text{MRG}}), \\[7pt]
 g(\mu_{ij}(\mb{s}_n,t)) & = & \sum_{k=1}^K \left\{\zeta_{k,n} \sin\left(\frac{2\pi k t}{\delta}\right)+\zeta'_{k,n} \cos\left(\frac{2\pi k t}{\delta}\right)\right\}.
 \end{array}
\end{equation}
The posterior distribution of $\bs{\theta}_{\text{time},n}=\{\zeta_{k;n}, \zeta'_{k;n}, k=1,\ldots, K\}$ is then obtained, and in the following steps these parameters are considered fixed at their posterior mean. This conditional approach has in general a small impact on the overall assessment of the uncertainty, as theoretical results \citep{edw20} have shown asymptotic correct results, and numerical results suggest an overall small impact or error propagation \citep{cas17}. Inference can be performed independently on multiple cores on a laptop or a workstation. The fitted values based on the model inference described above for two selected locations in Figure \ref{fig:data} are represented by black dashed line in Figure \ref{fig:wind}.

\subsection{Step 2: Covariates and Spatial Effects}

Once the temporal parameters are estimated, the parameters $\bs{\theta}_{\text{MRG}}, \mb{\beta}_{\text{ALT}}, \bs{\theta}_{\text{space}}$ associated to the marginal model, the covariate effect parameters and the spatial dependence in \eqref{eq:baye}, respectively, must be estimated. 

Inference in this step is especially challenging given the large number of spatial locations, which imply a large covariance matrix in \eqref{eq:fanova}, and hence challenges in storing it and performing linear algebra operations. If the diffusion-reaction SPDE \eqref{eq:baye}, as well as its generalization \eqref{eq:nonstat}, have a smoothness parameter fixed to an integer number (in this work, we assume $\nu_{\ell}=1$), then the underlying process can be shown to have the Markov property \citep{lindgren}. This implies that the continuous solution can be conveniently discretized via finite volumes with a discrete Gaussian Markov Random Field with a sparse precision matrix, hence leading to fast and affordable inference.

The SPDE in \eqref{eq:nonstat} is solved by dividing the domain into a triangulation, and then approximating the continuous solution by a piecewise linear function for each triangle, with the quality of the approximation being determined by the size of the triangulation \citep{lindgren}. Formally, the finite element approximation of the solution to the SPDE is
\[
\beta_{\ell}(\mb{s})\approx\sum_{k=1}^T \psi_k(\mb{s})w_k,
\]
where ${\psi_k}(\mb{s})$ are piecewise linear basis functions that are equal to one at vertex $k$, linearly decreasing to zero to nearby vertices, and zero everywhere else. The weights $w_k$ are Gaussian distributed, and $k=1,\ldots, T$ where $T$ is the number of vertices in the triangulation. We define three $T \times T$ matrices as:
\[
\begin{array}{rcl}
 \mb{C}=\text{diag}(\mb{C}_{ii})&,&\mb{C}_{ii} = \langle \psi_i,1 \rangle, \\[7pt] 
 \bs{G}_{ij} & = & \langle \nabla\psi_i, \nabla\psi_j \rangle, \\[7pt] 
 \bs{K} & = & \kappa_\ell^2\bs{C} +\bs{G},
 \end{array}
\]
where $\langle \cdot \rangle$ is the inner product and $\nabla$ is the gradient. The sparse precision matrix of the joint distribution of the weights in the case of a stationary SPDE \eqref{eq:spde} can be written as \citep{lindgren}:
\begin{equation} \label{eq:sol}
 \bs{Q} =  \bs{K}\bs{C}^{-1}\bs{K} = \tau_\ell^2\left( \kappa_\ell^4\bs{C}+ 2\kappa_\ell^2\bs{G}+\bs{G}\bs{C}^{-1}\bs{G}\right).
\end{equation}

If we further define $\bs{T}$ and $\bs{K}$ as diagonal matrices, where $\bs{T}=\text{diag}(\tau(\mb{s}))$ and $\bs{K}=\text{diag}(\kappa(\mb{s}))$, then the precision matrix for the nonstationary SPDE \eqref{eq:nonstat} can be written as:
\[
 \bs{Q} =  \bs{T}(\bs{K}^2\bs{C}\bs{K}^2+\bs{K}\bs{G}+\bs{G}^\top\bs{K}+\bs{G}\bs{C}^{-1}\bs{G})\bs{T}.
\]

\subsection{Bayesian Inference for Latent Gaussian Models}

In order to further ease the computational burden, Bayesian inference for the proposed model \eqref{eq:baye} will be performed using the INLA approximation \citep{rue}, a deterministic method for fast approximation of high dimensional integrals in Bayesian inference. A comprehensive review of INLA can be found in the works of \cite{rue} and \cite{bakka}. 

The INLA approach assumes a latent Gaussian model, of which our FANOVA model \eqref{eq:fanova} is a particular case, and throughout this section we omit the subscripts indicating the input and the temporal component.  We therefore assume we have a vector $\bs{Y}=(Y(\mb{s}_1), \ldots, Y(\mb{s}_n))^\top$ sampled at locations $\mb{s}_1, \ldots, \mb{s}_n$ whose marginal distribution is from the exponential family, possibly described by the hyperparameter vector $\bs{\theta}$. We assume that, conditional on a latent spatial field $\bs{x}$, the observations are marginally independent:
\[
\pi(\bs{y}|\bs{x},\bs{\theta})=\prod_{i=1}^n \pi(y(\mb{s}_i)|x(\mb{s}_i)),
\]
where $\bs{x}=(x(\mb{s}_1),\ldots, x(\mb{s}_n)^\top$ is a zero mean Gaussian field modeled with the SPDE approach in the previous section, with precision matrix $\bs{Q}(\bs{\theta})$. Under this model, the joint posterior distribution of the latent effects and hyperparameters can be written as:
\[
\begin{array}{rcl}
\pi(\bs{x},\bs{\theta}|\bs{y}) &\propto& \pi(\bs{\theta})\pi(\bs{x}|\bs{\theta})\prod_{i=1}^n\pi(y(\mb{s}_i)|x(\mb{s}_i),\bs{\theta}) \\[7pt]
& \propto & \pi(\bs{\theta})|\bs{Q}(\bs{\theta})|^{1/2}\exp\left\{-\frac{1}{2}\bs{x}^\top\bs{Q}(\bs{\theta})\bs{x}\right\}\prod_{i=1}^n\pi(y(\mb{s}_i)|x(\mb{s}_i),\bs{\theta}),
 \end{array}
\]
where $|\bs{Q}(\bs{\theta})|$ is the determinant of the precision matrix. In order to perform inference, it is of interest to obtain 1) $\pi(\bs{x}|\bs{y})$ from which all the marginal distributions $\pi(x(\mb{s}_i)|\bs{y})$ can be obtained; and 2) $\pi(\theta_j|\bs{y})$, the marginal posterior distributions of the hyperparameters. Both distributions can be obtained from the following integrals:
\[
\begin{array}{rcl}
\pi(x(\mb{s}_i)|\bs{y}) &=& \int \pi(x(\mb{s}_i)|\bs{\theta},\bs{y})\pi(\bs{\theta}|\bs{y})\mathrm{d}\bs{\theta}\\[7pt]
\pi(\theta_j|\bs{y})&=&\int \pi(\bs{\theta}|\bs{y})\mathrm{d}\theta_{-j}.
 \end{array}
\]
INLA predicates a Laplace approximation of the hyperparameter posterior $\tilde{\pi}(\bs{\theta}|\bs{y})$ using a Gaussian distribution. Such approximation would then allow us to obtain the posterior marginals of the latent parameter $x_i$ as:
\[
\tilde{\pi}(x(\mb{s}_i)|\bs{y})=\sum_{k}\tilde{\pi}(x(\mb{s}_i)|\bs{\theta}_k,\bs{y})\times \tilde{\pi}(\bs{\theta}_k|\bs{y})\times \Delta_k
\]
where $\Delta_k$ are the weights associated with a vector $\bs{\theta}_k$ of hyperparameters in a grid.  

To summarize, fast Bayesian inference in this work is obtained by exploiting 1) the latent Gaussian structure of the model, which allows an efficient deterministic Laplace approximation of the integrals determining the posterior distributions; and 2) a spatial model for the latent Gaussian field obtained through a solution of a SPDE, which results in a sparse precision matrix and allows us to take advantage of computationally efficient sparse linear algebra operations. In the supplementary material, we compare this inferential approach to a standard Markov Chain Monte Carlo (MCMC) inference for this Bayesian model, and we experienced that INLA was more than twice as fast, while delivering approximately the same distributions, see Figure S3. 

\section{Simulation Studies}\label{sec:sim}
In this section, we compare the ability to capture the spatial structure of the proposed nonstationary approach in \eqref{eq:nonstat}, which we denote as NSTAT, against a standard ANOVA with no spatial dependence (denoted as IND), as well as a spatial model with stationary SPDE in \eqref{eq:spde} (denoted as STAT). We show results for two cases: a Gaussian and a Bernoulli latent model. In Section \ref{sec:simdes} we outline the simulation design, in Section \ref{sec:metrics} we introduce the metrics used to compare the models, while in Sections \ref{sec:gaures} and \ref{sec:ngaures}, we show the results for the Gaussian and Bernoulli case, respectively. 

\subsection{Design}\label{sec:simdes}
The simulation is designed as a one-factor model on a 15 $\times$ 15 regular grid, for a total of $N=225$ locations. We assume a simplified setting of the WRF simulations with a single two-level factor, each with two replicates and with a deterministic structure over the spatial domain. Specifically, we denote our simulated response variable at location $\mb{s}_n, n=1,\ldots, N$, factor level $i \in \{0,1\}$ and replicate $j \in \{1,2 \}$ as
\begin{equation}\label{eq:sim}
\begin{array}{lll}
Y_{ij}(\mb{s}_n) & \sim & h(\mu_{i}(\mb{s}_n),\bs{\theta}_\text{MRG}),\\
g(\mu_{i}(\mb{s}_n)) & = & \beta_0+i\beta_1 I(\mb{s}_n \in \mathcal{D})+\epsilon_{ij}(\mb{s}_n),
\end{array}
\end{equation}

\begin{sloppypar}
\noindent where $Y_{ij}(\mb{s}_n)$ follows a distribution $h$ from the exponential family, $g(\cdot)$ is the link function, the error $\epsilon_{ij}(\mb{s}_n) \sim \mathcal{N}(0,\sigma^2)$ is independent and identically distributed across level $i$, replicate $j$ and location $\mb{s}_n$. We choose a constant nonzero value of $\beta_1$ for a sub-domain $\mathcal{D}$ and zero outside of it. We choose four shapes for $\mathcal{D}$: square, zigzag, bar and U (see Figure S3). We report here only the results for the square, while the three other shapes are shown in the supplementary material. We evaluate 1) a Gaussian model with identity link and $\beta_1= (100, 10, 2, 1.5, 1.2, 1, 0.5, 0.3, 0.1)$, $\beta_0=0$ and $\sigma=1$; and 2) a Bernoulli model with logit link with $(\beta_0, \beta_1)= ((-5, 100), (-5, 10) , (-1, 2) , (-1, 1), (-1, 0.5) , (-1, 0.1) )$ and $\sigma=0.1$. 
\end{sloppypar}

For each simulation, we fit the models assuming
\begin{equation}\label{eq:simmod}
g(\mu_{i}(\mb{s}_n))=\beta_0+i\beta_1(\mb{s}_n)+\epsilon_{ij}(\mb{s}_n),
\end{equation}
where $\bs{\beta}_1=(\beta_1(\mb{s}_1),\ldots, \beta_1(\mb{s}_N))^\top$ is a Gaussian random field. IND is a standard ANOVA assuming independence across all the locations, STAT is the stationary SPDE model \eqref{eq:spde} and NSTAT is the nonstationary SPDE model obtained solving \eqref{eq:nonstat}. A total of $n_{\text{sim}}=1,000$ simulations are performed for each shape and parameter combination.

\subsection{Metrics}\label{sec:metrics}

We assess the accuracy of the coefficient estimation using three metrics, two focused on the point estimates, and one on the spatial smoothness of the estimated latent process. The first one is the proportion of 95\% credibility intervals across the $n_{\text{sim}}$ simulations containing zero for each location. Ideally, this proportion should be close to 0\% inside $\mathcal{D}$ and 100\% outside of it. In the Gaussian case, the second metric is the proportion of 95\% credibility intervals that contains the true value of $\mu_i(\mb{s}_n)$ across the entire domain. The extent to which this proportion is closer to the nominal 95\% level is a measure of a correct uncertainty quantification, and better results uniformly across models are expected as the signal to noise ratio of the true model \eqref{eq:sim} increases. In the case of the Bernoulli model, for the second metric we instead use the receiver operating characteristic (ROC) curve, obtained by plotting the true positive rate against the false positive rate at different thresholds. In particular, we focus on the area under the curve (AUC) of the ROC curve, which is between zero and one, with the latter representing perfect prediction.

Finally, we evaluate the model's ability to capture the smoothness of the posterior mean of the latent process by assessing the discretized gradient averaged across the four directions:
\begin{equation}\label{eq:diff}
D_n=\frac{1}{|\partial \mb{s}_n|}\sum_{k \in \partial \mb{s}_n} \left\{\hat{\beta}(\mb{s}_n)-\hat{\beta}(\mb{s}_k)\right\},
\end{equation}
where $\partial \mb{s}_n$ is the set of the nearest neighbors of $\mb{s}_n$, and $|\partial \mb{s}_n|$ is its cardinality, which is 4, 3, or 2 depending on whether the point is in the middle, at edge or on the corner of the grid, respectively. The spatially varying estimate $\hat{\beta}(\mb{s}_n)$ represents the posterior mean. Since the true underlying process as specified in \eqref{eq:sim} is nonzero and constant across $\mathcal{D}$, a measure of performance of the models is the extent to which $\hat{\beta}(\mb{s}_n)$ have a small gradient $D_n$ across all points in the domain but at the boundary of $\mathcal{D}$. 

\subsection{Results: Gaussian Model}\label{sec:gaures}

\begin{figure}[ht!]
\includegraphics[scale=0.41]{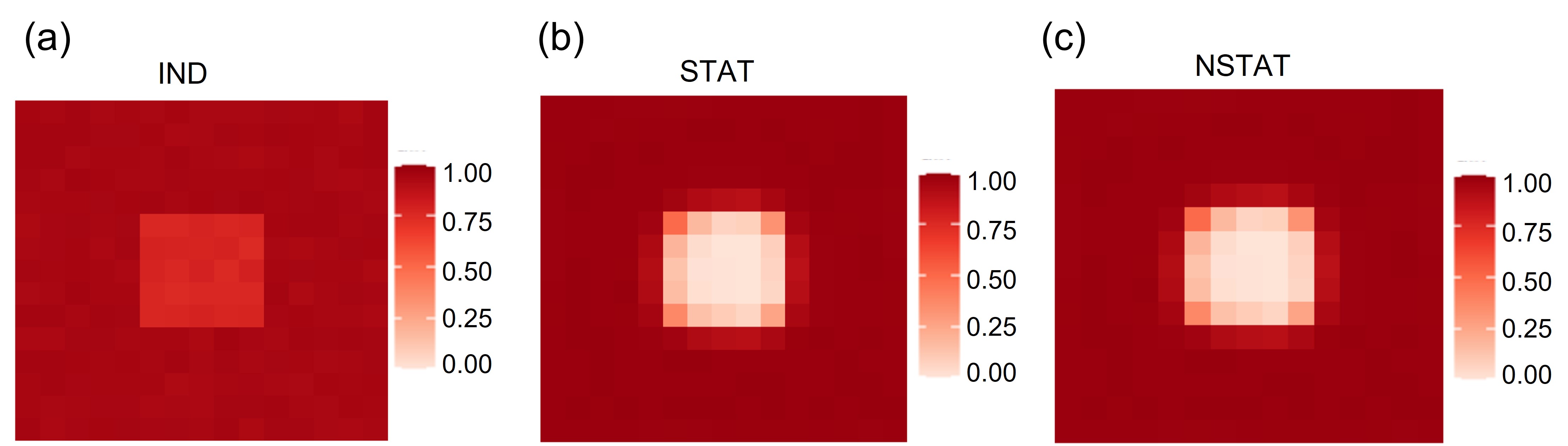}
\caption{Proportion of 95\% credibility intervals containing zero for the square design. We compare the models assuming spatial (a) independence (b) stationarity and (c) nonstationarity.}
\label{fig:prop}
\end{figure}

In Figure \ref{fig:prop} we show the map of the proportion of 95\% credibility intervals containing zero according to the three models for $(\beta_0,\beta_1)=(0, 2)$. The IND model in panel (a) is able to capture the change in value within the square, but the lack of spatial dependence results in noisy estimates both inside and outside the square, as spatially closer estimates cannot borrow strength from each other. As a result, a substantial number of simulations will flag zero values inside the square, where the expected percentage is 0\%. STAT and NSTAT in panels (b) and (c) both account for spatial dependence, hence they avoid noisy spatial estimates and are able to capture a change in pattern at the border of the square, as well as a nonzero signal inside of it.

\begin{figure}[ht!]
\includegraphics[scale=0.45]{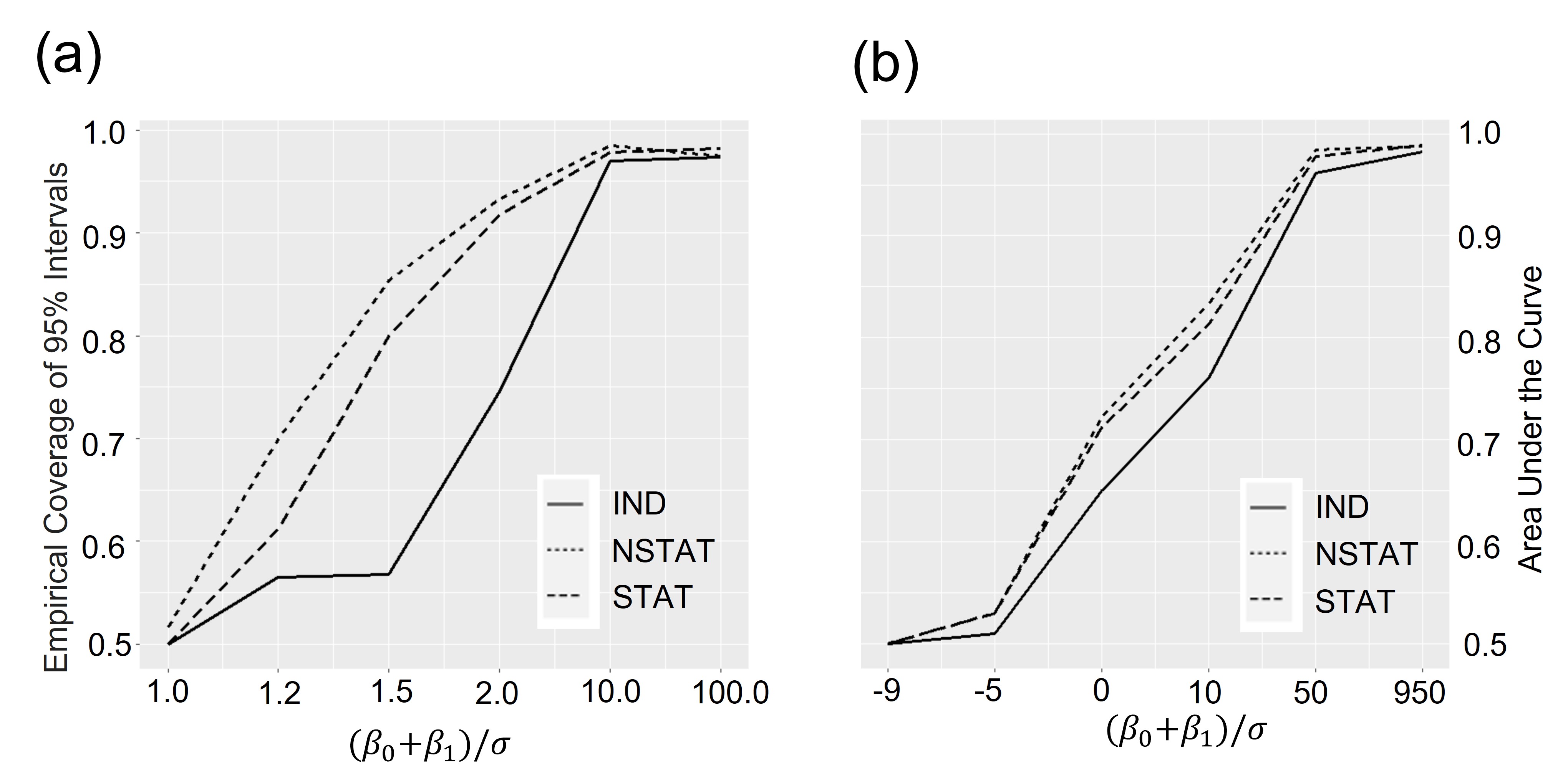}
\caption{(a) Proportion of 95\% credibility intervals (in the Gaussian case) and (b) area under the curve (in the Bernoulli case) across the $n_{\text{nsim}}=1,000$ simulations containing the true values against $(\beta_0+\beta_1)/\sigma$ in the square design. The $x$ axis in the log scale.}
\label{fig:propchange}
\end{figure}

In Figure \ref{fig:propchange}(a) we show the second metric, i.e., the empirical coverage of the 95\% credibility intervals (averaged across the domain) as a function of the signal to noise ratio. As expected, all models show an improvement for a stronger signal, as the characterization of the latent process and its uncertainty quantification is more apparent from the data. The IND model has consistently the worst performance and is able to recover a coverage close to the nominal level only for a very strong signal. The presence of a spatial effect for STAT and NSTAT allows one instead to have a considerably closer coverage for weaker signals, and consistently better results for NSTAT, given its ability to better capture the complex structure of the simulated process. Figures S4, S5 and S6 show similar results for the zigzag, bar and U shape, respectively, while Figure S7 shows similar results for all four shapes when we instead compare the predictive accuracy by means of the mean squared error. 

\begin{figure}[ht!]
\includegraphics[scale=0.6]{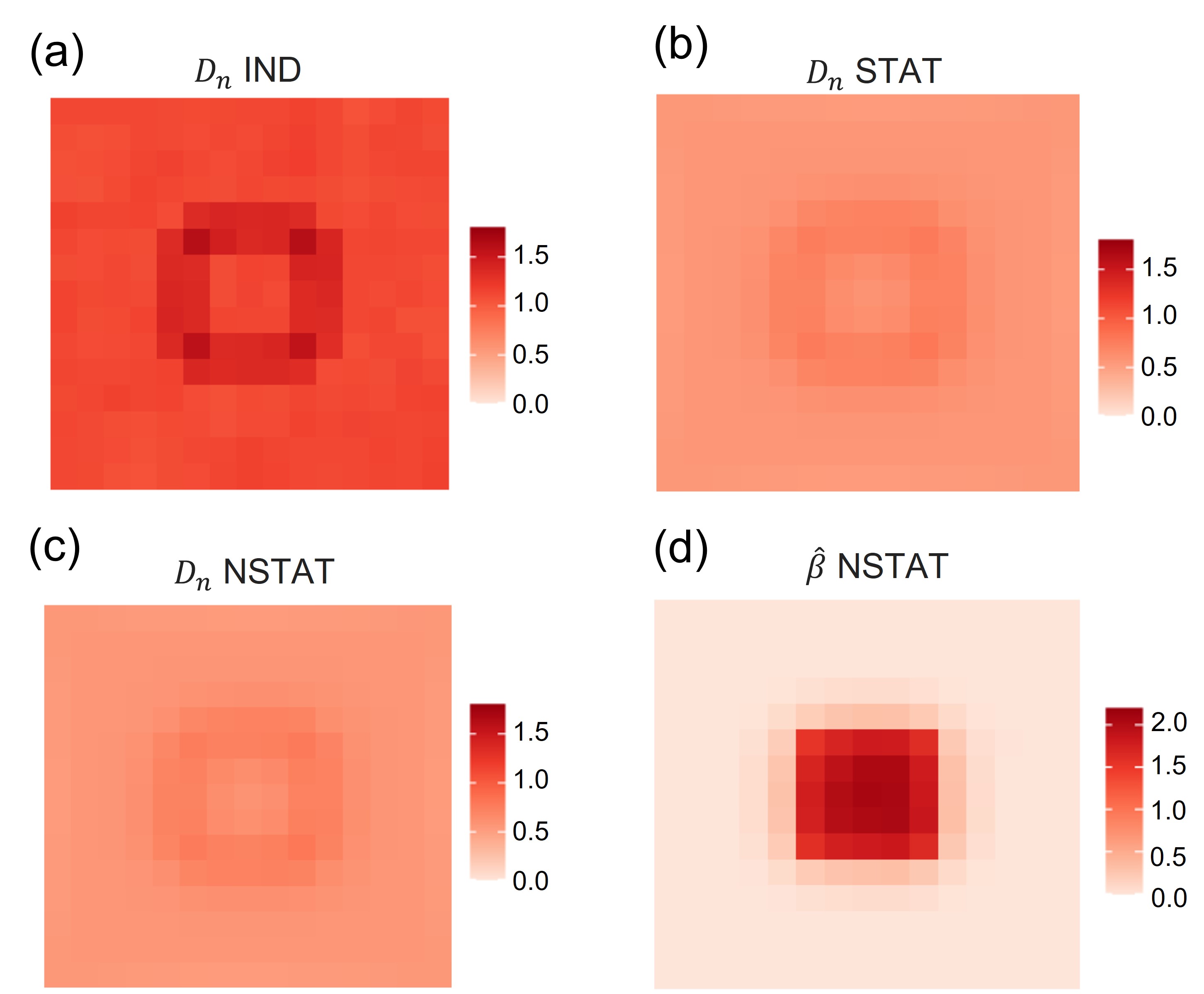}
\caption{Map of the posterior mean discretized gradient $D_n$ as defined in \eqref{eq:diff}, averaged across simulation from model \eqref{eq:sim} with $\beta=2$ and a square design. Results are shown for the model \eqref{eq:simmod} assuming (a) IND (median value across locations 1.13, IRQ 0.029), (b) STAT (0.14,0.06), (c) NSTAT (0.07,0.05) (d) The map of estimated $\beta$ values for NSTAT.}
\label{fig:betadiff}
\end{figure}

Figure \ref{fig:betadiff} shows a comparison of IND, STAT and NSTAT in terms of the discretized gradient $D_n$. In the case of IND in panel (a), the overall square shape is estimated, but the lack of spatial dependence results in non-smooth estimators and occasional local discontinuities, with an associated nonzero gradient outside the boundaries of $\mathcal{D}$. The  STAT model in panel (b) results instead in a smooth process and smaller gradients outside the square. The lack of flexibility in capturing local changes in the spatial structure however implies an overall lack of ability in capturing the spatial patterns on the square, and in particular the identification of a constant value inside it. The nonstationary model NSTAT in panel (c) is instead able to better capture the spatial structure both outside, inside and at the border of the square. Figures S7, S8 and S9 show similar results for the zigzag, bar and U shape, respectively. 

\subsection{Results: Bernoulli Model}\label{sec:ngaures}

In Figure \ref{fig:propchange}(b), we show the AUC as a function of the signal to noise ratio. For very weak signal, the three models have AUC close to 0.5, which corresponds to the predictive performance of a random guess. As the signal increases, however, all models increase their predictive ability. As in the Gaussian case, IND performs suboptimally given its inability to borrow strength across locations in space, STAT has better predictions, and NSTAT further improves them. Figure S10 shows a comparison of $D_n$ for IND, STAT and NSTAT with the Bernoulli simulations. As in the Gaussian case, the nonstationary model outperforms the other two by better capturing the spatial structure.

\section{Application}\label{sec:app}

In this section, we use the proposed functional ANOVA approach to estimate how the PBL and resolution locally affect the wind characteristics in Saudi Arabia. In Section \ref{sec:extra} we detail our approach to extrapolate the wind to turbine hub height. In Section \ref{sec:windeng} we evaluate the sensitivity of the ensemble runs with respect to PBL and resolution for both wind speed and wind energy. Section \ref{sec:thresh} further discusses a non-Gaussian sensitivity analysis of threshold exceedances. 

\subsection{Wind Speed Extrapolation}\label{sec:extra}

In this section we denoted by $W(\mb{s},h,t)$ the wind speed at location $\mb{s}$, height $h$ and time $t$. Since wind turbines are characterized by a hub height between 80m to 110m, extrapolation is necessary. Numerous literature studies have proposed models to characterize the vertical profile of wind, see \cite{gua19} for a recent review. A commonly used approach is to assume that the wind speed is directly proportional to the height through a power coefficient. The \textit{power law} stipulates that:
\begin{equation} \label{eq:power}
\begin{array}{rcl}
 W(\mb{s}_n,h_k,t) & = & W(\mb{s}_n,h_r,t)\left(\frac{h_k}{h_r}\right)^{\alpha_{n,t}}e^{\eta(\mb{s}_n,t)},\\[7pt]
 \eta(\mb{s}_n,t) & \sim &  \mathcal{N}(0,\sigma^2_{n,t}),
\end{array}
\end{equation}
where $h_k$ is the hub height to which we want to extrapolate, and $h_r$ is the reference height at which wind speed data are available. The \textit{shear coefficient} $\alpha_{n,t}$ controls the magnitude of increase in the mean wind speed as the height increases \citep{gua19}. In the absence of detailed meteorological observations, the shear coefficient is often assumed to be constant in space and time: $\alpha_{n,t}=\frac{1}{7}$. This choice is based on the assumption of a flat surface and neutral atmospheric conditions \citep{pat78}. However, this simplification has been shown to be largely inappropriate in Saudi Arabia \citep{cri21}. To estimate the shear coefficient, our WRF simulations provide the wind speed vertical profile at six levels, at approximately equally spaced heights from 20m to about 110m, in addition to the wind speed at the reference height of 10m. A simple log regression can be applied to estimate $\alpha_{n,t}$ using the six levels of wind speed for each of the four WRF runs. An example of the estimated shear coefficients, their standard deviation and the coefficients of determination $R^2$ for the run MYJ-6km are displayed in Figure S11. In the mountainous Hijaz area (see Figure S1) in the west of Saudi Arabia, we observe very small estimated $\alpha_{n,t}$ values, even negative at some locations, indicating that the wind speed is on average lower at higher altitudes; this is an uncommon situation associated with large standard deviations and small $R^2$ values. These results suggest that the power law may not be suitable for some geographic locations, as discussed in previous studies \citep{gua19,cri21}. However, a previous study highlighted that the installation of a wind turbine over the complex terrain of the Hijaz region is not cost effective \citep{gia20}. Therefore, the negative estimates in these areas are not a cause of concern in this study.

The wind speed at hub height is then converted to wind energy using turbine-specific \textit{power curves} that transform wind speed to power. A power curve assumes a value of zero until a minimum cut-in speed is reached and the blades start rotating; then, it keeps increasing with stronger wind speeds. Finally, the power curve reaches a cut-off speed and remains constant thereafter. We chose the turbine makes and models that were identified as the most cost effective for each location by \cite{gia20}, see Figure S12. To compute the total wind energy that can be generated in each grid cell of our domain, we multiplied the power of a single turbine by the number of turbines that can fit in a specific grid cell. This number depends on the length of the turbine blades, as sufficient spacing must be provided to prevent excessive local turbulence. 

\subsection{Sensitivity of Wind Speed and Energy}\label{sec:windeng}

We applied the FANOVA model \eqref{eq:baye} with a Gaussian marginal distribution and an identity link to both ground wind speed and energy and assess the local sensitivity. A triangulation comprising $T=53,387$ triangles across the domain was chosen for the SPDE approach, see Figure S13 for a plot of the triangulation in a subdomain. We chose $K=3$ harmonics for the temporal component, as indicated by the model selection in Figures S14 and S15. The temporal trend can resolve the dynamics of the data, Figure S16 shows the boxplot of the skewness and excess kurtosis from the temporal residuals of the model for both wind speed and energy. Overall, the vast majority of locations show a symmetric distribution with no excess kurtosis, hence lending support for the Gaussian assumption. Figure S17 shows the posterior distribution of the non-spatial altitude effect $\beta_{\text{ALT}}$ and the nugget effect $\sigma^2$ in the FANOVA model \eqref{eq:baye}. Consistently with the diagnostics in Figure S2, $\beta_{\text{ALT}}$ is positive with very high probability, hence wind increases with altitude.

\begin{figure}[ht!]
\includegraphics[scale=0.52]{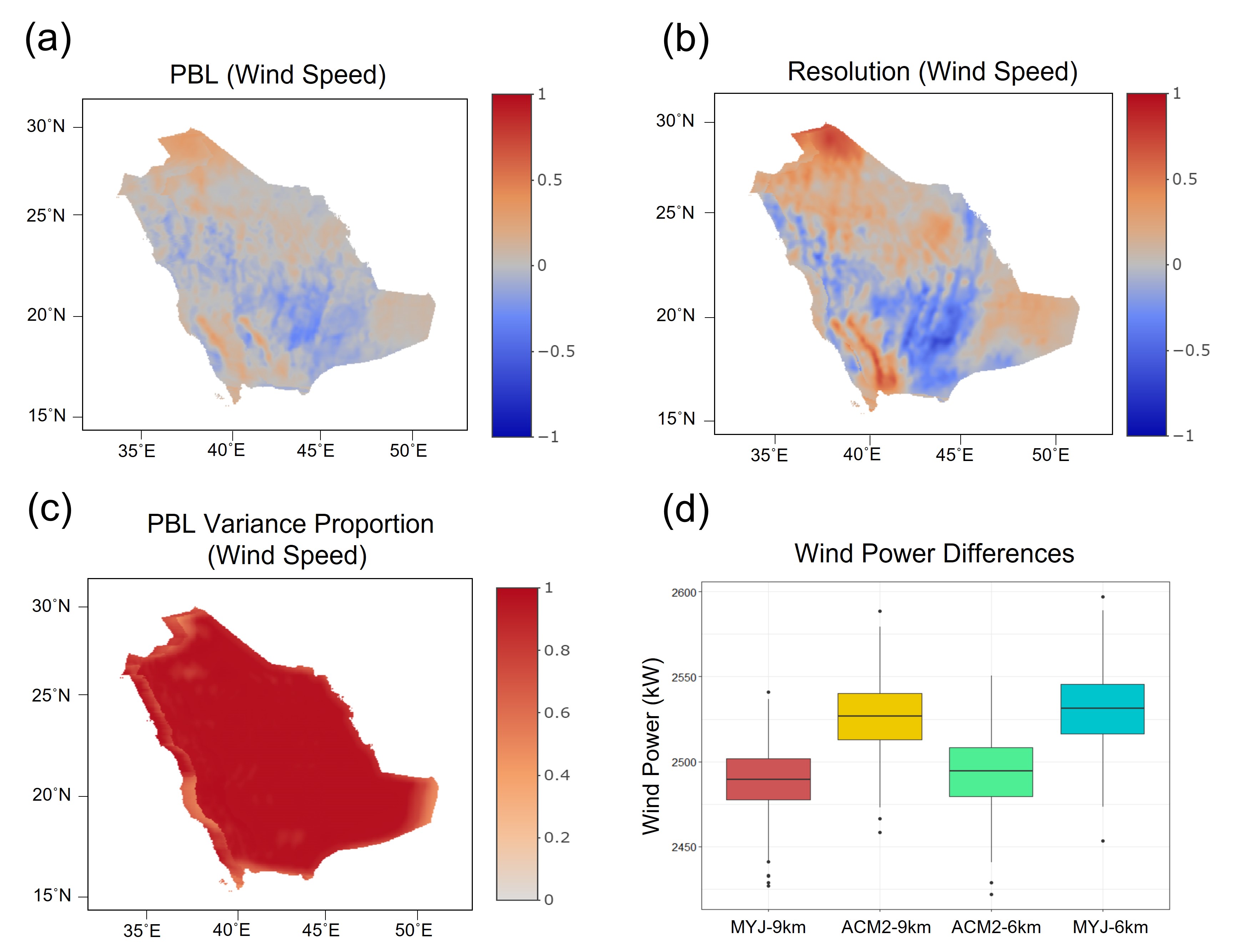}
\caption{Posterior mean of the (a) PBL and (b) resolution coefficient of surface wind speed. Panel (c) shows the proportion of posterior variance the from PBL coefficient against that of resolution. Panel (d) shows the distribution of wind power for the 75 optimum wind farm locations in \cite{gia20} by sampling from the parameters' posterior distribution.}
\label{fig:result}
\end{figure}

Figure \ref{fig:result} shows the estimated posterior mean of $\bs{\beta}_{\text{PBL}}$ and $\bs{\beta}_{\text{RES}}$ using the FANOVA model described in equation \eqref{eq:baye} for wind speed, and Figure S18 shows the corresponding results for wind energy. Figure \ref{fig:result}(a) indicates that the ACM2 PBL scheme tends to simulate higher wind speed than the MYJ scheme over the northern portion of the domain and over the areas of complex terrain in the southwest area of Saudi Arabia. The difference between the two schemes is small over large areas of the domain, and MYJ dominates in central Saudi Arabia. Figure \ref{fig:result}(b) shows similar spatial patterns for the simulation resolution, suggesting that both the higher resolution and the choice of the ACM2 PBL scheme result in higher winds in the same regions, some of which are characterized by complex topographic features that indeed require higher resolution and/or complex physics schemes to capture the flow patterns in these complex terrains accurately. This is expected as ACM2 combines the nonlocal and local turbulence scheme for unstable and stable conditions, respectively, unlike MYJ which is strictly local. These results agree with prior studies showing higher accuracy of ACM2 than MYJ, also over complex terrain \citep{siuta17,Hu10}. The posterior means of the PBL and resolution for wind energy are shown in Figures S18(a) and (b), respectively. The spatial patterns are similar to those of wind speed, with a generally higher posterior mean (and hence a stronger effect), which is a feature attributable to the vertical extrapolation described in the previous section, as wind at hub height is generally higher. We also show in Figure \ref{fig:result}(c) the ratio between the posterior variance of $\bs{\beta}_{\text{PBL}}$ against the total posterior variance from both effects (see Figure S18(c) for the corresponding results for wind energy), and it is readily apparent how PBL is more variable than the resolution for almost all points in the domain, except noticeably in the western coast, where a sharp transition between the Hijiaz mountain range and the sea can be better characterized by an increased resolution. 

For the 75 locations indicated by \cite{gia20} as the most cost-effective for building wind farms, we can use the proposed model to assess the differences in total monthly wind power output across ensemble members. Indeed, we fix the parameters to their posterior mean, simulate 500 realizations of surface wind speed, extrapolate them and compute the power output according to the power curves. Figure \ref{fig:result}(d) shows the boxplot of the wind power distribution across the simulations for all four ensemble members. The MYJ-9km simulation would result in an average output of 2488.5 kW, whereas ACM2-9km would increase this on average by 11.95 kW (0.4\%), MYJ-6km by 22.94 kW (0.9\%), and ACM2-6km by 34.89 kW (1.4\%). The FANOVA therefore lends further support for the siting work in \cite{gia20}, which was performed with MYJ-6km (the simulation with closer validation metric with respect to some ground observations), by showing an overall robustness of the energy output with respect to both PBL and resolution. 

The same analysis was performed using a standard ANOVA (i.e., assuming spatial independence) and a stationary FANOVA, as shown in Figure S19. In the case of ANOVA, in order to evaluate the impact of resolution, spatial interpolation of wind on the same is necessary, so we upscale the 6$\times$6km simulations to the 9$\times$9km grid of the other simulations in the ensemble. The stationary FANOVA model does not require any upscaling. The standard ANOVA suggests that ACM2 PBL scheme tends to simulate slower wind speed along the coast line (dark blue color in the map) and that resolution has smaller impact than the nonstationary FANOVA, with near zero estimates at the majority of the locations. The stationary FANOVA results show that ACM2 PBL scheme tends to simulate high wind speeds in the northern region. With regard to resolution, we observe non-zero estimates at most of the locations. We can therefore conclude that the use of a nonstationary FANOVA emphasizes the role of the adopted resolution and PBL parameterization, especially in the context of complex terrains, which is consistent with findings from previous physics-based modeling studies.

We also assessed the smoothness of the latent spatial factors using the discrete Laplacian for the ANOVA model, as well as the FANOVA stationary and nonstationary model. For wind speed, a decrease of 33\% (31\% for wind energy) of the averaged discrete Laplacian was observed using FANOVA with stationary assumption compared to the ANOVA model, and a further decrease of 25\% (18\%) was achieved using the nonstationary FANOVA model. Similar improvements in the smoothness were observed for wind energy data.

\subsection{Sensitivity of Threshold Exceedances}\label{sec:thresh}

\begin{figure}[ht!]
\includegraphics[scale=0.57]{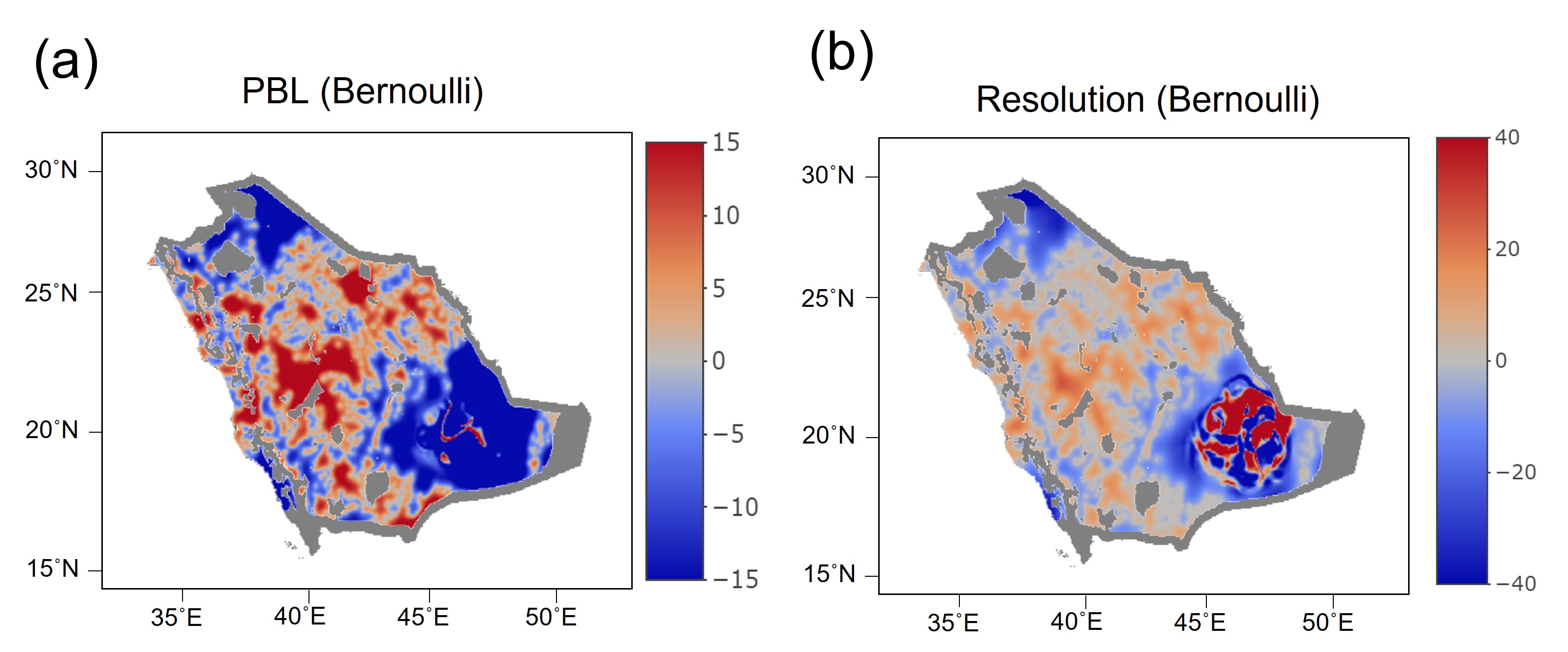}
\caption{Posterior mean of the (a) PBL and (b) resolution coefficient of the Bernoulli model.}
\label{fig:nongaus}
\end{figure}

To further assess the sensitivity of wind energy with respect to the PBL and resolution, we considered the threshold exceedance of wind power with respect to half of the maximum power output according to the power curve. The response variable $\mb{Y}_{ij}(t)$ therefore follows a Bernoulli distribution, with a logit link function $g(\cdot)$ in \eqref{eq:baye}. The temporal parameters $\bs{f}^{(T)}(t)$ in the FANOVA model \eqref{eq:baye} were estimated independently at each location using Bayesian logistic regression. Figure S20 shows the binary values and fitted values of the Bernoulli model for the P and M locations as indicated in Figure \ref{fig:data}. The posterior means of the temporal effects were assumed to be fixed and the spatial effect was estimated. Figures \ref{fig:nongaus}(a) and (b) show the posterior mean of $\bs{\beta}_{\text{PBL}}$ and $\bs{\beta}_{\text{RES}}$, respectively. For the PBL coefficient, large negative values are observed in the Rub' al Khali area, indicated by the dark blue color on the map. This implies that on average the non-local ACM2 PBL scheme yields a decrease in the threshold exceedances. For the resolution coefficient, large positive values in the posterior mean are apparent in the same area as for the PBL. In other words, in the Rub' al Khali region, the choice of 6km resolution tends to increase the odds of threshold exceedance. In the central Saudi Arabia and Hijaz region, both the ACM2 PBL scheme and 6km resolution tend to increase the odds of threshold exceedance.

\section{Conclusion}\label{sec:con}

In this study, we propose a model-based local sensitivity analysis of a climate ensemble, with spatially varying latent fields modeled using a nonstationary SPDE. The proposed approach allows us to capture the spatial dependence for a complex dataset on a large domain while simultaneously accounting for non-Gaussianity. Furthermore, the assumption of a continuous underlying process allows us to perform the sensitivity study for simulations with different spatial resolutions without any \textit{ad hoc} upscaling. The simulation studies performed under a wide range of settings provide compelling evidence that the nonstationary model can capture more complex structures than a standard ANOVA or a stationary model. The proposed FANOVA approach is then used to provide insights about dependence of wind speed and energy for Saudi Arabia from PBL scheme and resolution.

The proposed model can be generalized to any sensitivity study consisting of spatio-temporally resolved ensembles, with categorical but also quantitative input. From a methodological point of view, the assumption of a spatially varying precision and range through SPDE and basis decomposition allows us to capture complex patterns of spatial dependence while simultaneously allowing for a valid process. Alternative domain-specific approaches are also possible: if the geography suggests changes in the dependence structure dictated by descriptors such as land/ocean, the SPDE operator can be modified to account for that with a more tailored spatial dependence structure \citep{fug20,hu21}. Additionally, for applications where several variables are of simultaneous interest (e.g., temperature, wind and precipitation), a multivariate approach can be proposed by relying on the sparse approximations of multivariate SPDEs \citep{hu13}, although the task of determining the dependence structure across both space and variables is currently limited by the dearth of sufficiently flexible models \citep{gen15}. 

From the perspective of the application of interest in this work, our results indicate that both wind speed and energy are sensitive to the resolution and PBL scheme, with a non-local scheme and high resolution generally resulting in higher winds speeds over complex terrain, consistently with previous geoscience literature in other world areas. Additionally, this work indicates that the current plan for building turbines in Saudi Arabia is robust with respect to the input chosen in the ensemble, as the final output estimates change by at most 1.4\% from the reference siting work \citep{gia20}. 

Finally, while the spatial sensitivity analysis offers scientific insights into the effect of the resolution and PBL schemes, the ultimate goal is to determine the extent to which these simulations offer an accurate representation of the true wind fields. The ensemble analyzed in this work has been validated with ground observations \citep{gia20}; however, the spatial coverage of the observational network was sparse and the diagnostic was limited to standard metrics in the geoscience literature. If a more comprehensive observational data set becomes available in the future, a more formal model-based approach can be proposed by assuming a true observational process in the FANOVA, thereby allowing both calibration and sensitivity analysis. 

\bibliographystyle{plainnat}
\bibliography{references}

\end{document}